\documentclass{PoS}

\usepackage{amsfonts,bm,amssymb,euscript,array,cite}
\usepackage[all,color]{xy}
\usepackage{relsize}
\usepackage{tikz}
\usepackage{url}
\usepackage{amsmath,amsthm}
\usepackage[]{graphicx}
\usepackage[T1]{fontenc}
\usepackage{framed}

\usepackage{empheq}

\makeatletter

\newcommand{\dd}{\mathrm{d}}

\newcommand{\w}{\wedge}
\newcommand{\bbm}{\left(\begin{matrix}}
\newcommand{\ebm}{\end{matrix}\right)}
\newcommand{\beq}{\begin{eqnarray}}
\newcommand{\eeq}{\end{eqnarray}}

\makeatother

\newcommand{\sfrac}[2]{{\textstyle\frac{#1}{#2}}}

\newcommand{\be}{\begin{equation}}
\newcommand{\ee}{\end{equation}}

\newcommand{\beqa}{\begin{eqnarray}}
\newcommand{\eeqa}{\end{eqnarray}} 
\def\nn{\nonumber} \def \bea{\begin{eqnarray}} \def\eea{\end{eqnarray}}

\newcommand{\barr}{\begin{array}}
\newcommand{\earr}{\end{array}}
\numberwithin{equation}{section}

\def\a{\alpha}  
 \def\g{\gamma} 
 \def\d{\delta} 

   \def\m{\mu}
 \def\o{\omega}   
\def\s{\sigma} \def\S{\Sigma}

\def\R{{\mathbb R}}   

 \def\one{\mbox{1 \kern-.59em {\rm l}}}

\def\bit{\begin{itemize}} \def\eit{\end{itemize}}

\def\({\left(} \def\){\right)}

\title{Non-isometric T-duality from gauged sigma models}

\ShortTitle{Non-isometric T-duality from gauged sigma models}

\author{\speaker{Athanasios Chatzistavrakidis}\thanks{Part of the work presented here was done in collaboration 
with Andreas Deser and Larisa Jonke \cite{Chatzistavrakidis:2015lga}. 
Further related collaboration with Thomas Strobl which contains more general results is acknowledged too \cite{strobl}.}\\
        Van Swinderen Institute for Particle Physics and Gravity, University of Groningen, \\ 
        Nijenborgh 4, 9747 AG Groningen, 
        The Netherlands\\
        E-mail: \email{a.chatzistavrakidis@rug.nl}}

\abstract{Local symmetries is one of the most successful themes in modern theoretical physics. 
Although they are usually associated to Lie algebras, a gradual increase of interest in more general 
situations where local symmetries are associated to groupoids and algebroids has taken place in 
recent years. On the other hand, dualities is another persistently interesting theme in modern 
physics. One of the most prominent examples is provided by target space duality in string theory. 
The latter, Abelian or not, is usually associated to the presence of isometries, which is however a very 
restrictive assumption. In this contribution we discuss some recent advances located at the 
intersection of the above two themes. Focusing on bosonic string sigma models we discuss certain gauged versions 
where (a) the invariance conditions on the background fields are much milder than the isometric case and 
(b) the gauge symmetry is generically associated to a Lie algebroid instead of just a Lie algebra. 
Furthermore we utilize such gauged sigma models to study the possibility of non-Abelian, non-isometric 
T-duality. 
}

\FullConference{Proceedings of the Corfu Summer Institute 2015 "School and Workshops on Elementary Particle Physics and Gravity"\\
		1-27 September 2015\\
		Corfu, Greece}

\begin{document}

\section{Motivation and Introduction}

The utmost importance of symmetry is modern theoretical physics is hard to overemphasize. The gauge principle, being the cornerstone of our theories describing fundamental interactions, has brought local 
symmetries to the forefront of attention since long ago. 
Conventionally, speaking of local symmetries what comes to mind 
is a Lie algebra and a set of Lie algebra-valued 1-forms called gauge fields that are introduced in the theory through minimal coupling. 
One question that comes to mind then is whether this is the most general 
setting, namely whether \emph{every} local symmetry is associated to 
such a scenario. It is known that this is not the case (see for example 
the inspiring exposition in Ref. \cite{weinsteinams}). 
Given the unquestionable success of Lie-algebra-based gauge symmetries 
it is certainly worth studying such more general scenarios, based for example on the notion of a Lie algebroid. Moreover, a second part of this 
question refers to minimal coupling and whether it is always enough to 
guarantee the existence of a gauge theory. The investigation of these 
questions in the context of two-dimensional sigma models is the 
first motivation for the work presented here.

Another important notion related to symmetry is duality. Dualities are 
everywhere in physics (see for example the exposition in 
Ref. \cite{Polchinski:2014mva}) and this is certainly the case in string theory, where they are very profound properties that teach us 
several conceptual lessons about the theory. One of the prime examples 
is T-duality which identifies string backgrounds associated to 
different target spaces. Conventionally, speaking of 
T-duality what comes to mind is a target space with one or more 
isometric directions.{\footnote{An interesting exception is Poisson-Lie 
T-duality \cite{Klimcik:1995ux,Sfetsos:1997pi}.}} This is however extremely restrictive. 
Certainly a randomly chosen string background has no isometries whatsoever. The immediate question is then whether this is the best we 
can do.  Recall that one approach to T-duality goes through an "intermediate" gauge theory which on-shell reduces to one or a dual 
background \cite{Buscher:1987sk,Rocek:1991ps}. Here we will study such gauge theories even 
when no isometries are available. The relation among gauged sigma models and T-duality is therefore our second general motivation.

Finally, there is a third motivation that is worth mentioning 
although it is not going to be addressed in this work. It regards 
the so-called non-geometric string backgrounds, which often originate 
from T-dualities and their precise understanding is still a 
programme under way. One certain lesson of recent studies on this topic 
is that their description, be it from the target space viewpoint or 
the world sheet one, requires generalized geometric concepts.
Here we are going to study gauged sigma models whose target 
space is essentially some generalized tangent bundle. 

It is remarkable that there is a common mathematical theme underlying 
\emph{all} three motivations above: the theory of groupoids and algebroids. 
Indeed, (i) behind every local symmetry there is a groupoid or an algebroid 
\cite{weinsteinams}, 
(ii) T-duality is mathematically described, under certain 
restrictions of course, as an isomorphism of 
(Courant) algebroids \cite{Cavalcanti:2011wu,Severa:2015hta}, and (iii) 
generalized geometry is partially yet crucially based on algebroid theory 
\cite{Gualtieri:2003dx}.

Based on the above motivations, we discuss a threefold generalization of the traditional picture of 
Lie algebra-based, minimally coupled 2D gauge theories with background fields satisfying strong invariance conditions. 
In particular, the first fold of the generalization is to replace the Lie algebra $\mathfrak{g}$ 
by a Lie algebroid{\footnote{A more general case where the algebroid 
is required to be just \emph{almost} Lie is discussed elsewhere \cite{strobl}.}} over the target space $M$. 
A Lie algebroid is an interesting mathematical structure that merges two very important notions for physics, namely 
algebras and vector bundles. At first approximation one could think of it as a generalization of an algebra such that 
the structure constants are not constant anymore but are instead replaced by \emph{structure functions}. One particularly illuminating 
way to see this is to study the famous Cartan problem, which is also relevant in many physical contexts, 
ranging from the vierbein formulation of gravity to string theory reductions. 
The problem can be stated as follows (see for example the excellent lectures notes \cite{crainic}, which we 
follow closely here):
given two sets of functions defined locally on $M=\mathbb{R}^n$, say $C^a_{bc}$ and $e^i_a$, where 
the indices range as $a=1,\dots, r$ and $i=1,\dots n$, we are asked to 
determine (i) a manifold $N$, (ii) a coframe $\{e^a\}$ on the manifold, namely a basis of its cotangent bundle $T^{\ast}N$, 
and (iii) a local coordinate system $X: N\to M=\mathbb{R}^n$, such that the following two equations hold:
\bea 
\dd e^a&=&-\sfrac 12 C^a_{bc}(X)e^b\w e^c~,\label{mc}
\\
\dd X^i&=&e^i_{a}(X)e^a~.
\eea
Asking when this problem has solutions, it is immediately observed that there are two necessary conditions obtained by taking 
the exterior derivatives of the above two equations,
\bea 
C^a_{e[b}C^e_{cd]}&=&e^i_{[b}\partial_iC^a_{cd]}~,\label{nc1}
\\
2e^j_{[b}\partial_je^i_{c]}&=&C^a_{bc}e^i_a~,\label{nc2}
\eea
where antisymmetrizations are taken with weight. A very familiar special case (e.g. from the vielbein formulation of 
gravity) corresponds to contant functions $C^a_{bc}$. Then the condition \eqref{nc1} becomes the usual Jacobi identity for 
Lie algebras, and the condition \eqref{nc2} becomes the closure of the algebra of the vectors $e_a$ dual to the coframe. 
Then one lands in the Lie algebra case and \eqref{mc} becomes the Maurer-Cartan equation. However, in general the 
functions $C^a_{bc}$ are not constant, in which case the solution to the Cartan problem is associated to a Lie algebroid. 
One can reach a good definition by using the necessary conditions \eqref{nc1} and \eqref{nc2}. The second condition 
indicates that we have to replace the Lie algebra $\mathfrak{g}$ by a vector bundle $L$, which is equipped 
with a bracket such that its sections $e_a$ close under it with the structure functions $C^a_{bc}$:
\be 
[e_a,e_b]_{L}=C^c_{ab}(X)e_c~.\label{bracket}
\ee
In order to associate this to \eqref{nc2}, an additional ingredient is a map from the vector bundle $L$ to the tangent 
bundle $TM$, i.e. a rule that assigns a vector field to every section of $L$, $\rho: L \to TM$. Then Eq. \eqref{nc2} 
simply says that this map is a homomorphism:
\be 
\eqref{nc2} \quad \Leftrightarrow \quad \rho([e_a,e_b]_L)=[\rho(e_a),\rho(e_b)]~.
\ee
Moreover, the other necessary condition acquires a simple explanation too:
\be 
\eqref{nc1} \quad \Leftrightarrow \quad [e_a,[e_b,e_c]_L]_L ~ + ~ \text{(cyclic permutations)}=0~;
\ee
in other words it becomes the Jacobi identity for the bracket on $L$, which is thus a Lie bracket. The above three 
ingredients, the vector bundle $L$ with a Lie bracket and a homomorphism $\rho$ to the tangent bundle of $M$, define 
a Lie algebroid. (The definition should be understood generally, not just for $M=\R^n$, which was used here as a clarifying example.)  Certainly Lie algebras are included in this definition, simply by taking $M$ to be just a point. 
For a list of examples we refer e.g. to \cite{crainic}.

It should now be fairly obvious what the first generalization amounts to. The idea is to replace the Lie algebra 
$\mathfrak{g}$ by a Lie algebroid
$L$ 
such that the ``generalized local symmetry'' is generated by vector fields that close under the Lie bracket with some 
$X$-dependent structure functions.

Let us now turn to 2D sigma models and explain what the second and third folds of the generalization are. 
Recall that the common bosonic sector of string theory includes the background fields $(g,B,\Phi)$, namely a target space metric, 
the Kalb-Ramond 2-form and the scalar dilaton. The theory is described by a non-linear sigma model whose source is a 
2D world sheet $\S$ and its target a manifold $M$, and the dynamical fields are the scalar components of the corresponding 
map $X=(X^i):\S\to M$. The corresponding action functional is 
 \bea\label{isomodel}
 S&=&\int_{\Sigma}\sfrac 12 g_{ij}(X)\dd X^i\wedge \ast \dd X^j
 +\int_{\hat\Sigma} \sfrac 16 H_{ijk}(X)\dd X^i\wedge\dd X^j\wedge\dd X^k
+\alpha'S_{\text{dilaton}}~, \eea
where $\ast$ is the Hodge operator on the world sheet ($\ast^2=\mp 1$ for Euclidean and Lorentzian signature respectively) 
and $H$ is a Wess-Zumino term, only locally exact such that $H=\dd B$, living on an open membrane world volume $\hat\S$ whose 
boundary is the world sheet $\S$. The dilaton coupling involves the world sheet curvature scalar and it is of next 
order in $\a'$. In this work it will therefore be ignored.

Suppose now that we are given a set of vector fields $\rho_a=\rho_a^i(X)\frac{\partial}{\partial X^i}$ which generate the following global symmetry
\be 
\d_{\epsilon}X^i=\rho^i_a(X)\epsilon^a~,\label{sym}
\ee
for rigid transformation parameters $\epsilon^a$. Then the action \eqref{isomodel} is invariant 
under this symmetry provided that the following two conditions hold:
\bea 
{\cal L}_{\rho_a}g&=&0~, \label{isog}
\\
\iota_{\rho_a}H&=&\dd\theta_a~,\label{isoH}
\eea
for some  1-form $\theta_a$. Eq. \eqref{isog} means that the vector fields $\rho_a$ are Killing, namely they generate isometries for the metric $g$. 
Then this global symmetry can be promoted to local one by introducing Lie algebra-valued 1-forms $A=(A^a)$ (gauge fields), allowing the 
parameters $\epsilon^a$ to depend on world sheet coordinates $\s^{\m}$ and also allowing the gauge fields to transform 
appropriately under such gauge transformations, in particular as follows:
\be 
\d_{\epsilon}A^a=\dd\epsilon^a+C^a_{bc}A^b\epsilon^c~,\label{AgaugeLie}
\ee 
with $C^a_{bc}$ here the structure constants of the Lie algebra.
This was studied in detail in \cite{Hull:1989jk,Hull:1990ms} and 
revisited recently in \cite{Plauschinn:2013wta,Plauschinn:2014nha,Bakas:2016nxt}. It can already be invoked from these 
works that minimal coupling is not enough to yield the correct gauged action functional. Thus the generalization to 
non-minimally coupled gauge fields is already present (and necessary) at this level and will be even more transparent 
in the body of this work. Moreover, already in the isometric case there are additional constraints that have to be satisfied 
in order for the $A$-extended action functional to be gauge invariant. This will also become transparent in the main text.

Most importantly, it should be stressed that the invariance conditions \eqref{isog} and \eqref{isoH} are extremely 
restrictive. Indeed an arbitrary choice of background fields is unlikely to satisfy them. Thus it would be 
important to be able to write down gauged action functionals that still permit \emph{local} symmetries of the type \eqref{sym} 
but without having to satisfy such restrictive conditions. Clearly this is not possible at the global level; however, 
starting from the inspiring work \cite{Kotov:2014iha,Kotov:2015nuz}, it turns out to be possible at the level of gauge symmetries. 
In other words, in an inversion of the usual spirit, we do not start with a global symmetry and then promote it to a 
local one; instead we search for local symmetries directly, given a candidate action functional that is likely to realize a desired 
symmetry. This is the third and most crucial generalization of the usual approach. We will see that upon an appropriate choice 
of gauge transformation for the 1-forms $A$, the invariance conditions are replaced by extremely milder ones, 
certainly such that 
\bea 
{\cal L}_{\rho_a}g &\ne & 0~,\label{vague1}
\\
\iota_{\rho_a}H-\dd\theta_a &\ne &0~,\label{vague2}
\eea
which allow 
for gauge theories without isometry. 

It should be stressed that the three above generalizations are to some extent independent. Indeed,  it is 
possible for example to drop isometry while keeping either minimal coupling or Lie algebra-valued gauge fields or both. 

Returning to the second motivation, that is T-duality, the generalization described above has direct consequences. One can 
now view the gauged action functionals as intermediate gauge theories in a Buscher-like approach to T-duality. 
In the present case the associated vector fields $\rho_a$ are neither required to be Abelian nor Killing. 
Thus, introducing Lagrange multipliers in the action functional so as to reduce the additional degrees of freedom 
introduced by the gauge fields $A$, it is possible to determine two different reduced models upon integrating out 
different fields in the action. Then these two resulting sigma models are in a sense dual to each other and correspond 
to backgrounds with different target space. At least at the classical level one can say that this is a version of 
non-Abelian and non-isometric T-duality. 

It is natural to worry whether the above possibility can really be realized in non-trivial examples. We will see that at the classical 
level this is indeed possible and we will discuss some particular cases. It is not known whether examples surviving at 
the quantum level exist, but this is an interesting open question which we leave for future investigations. 

\section{Beyond the standard gauging in 2D sigma models} 

As explained in the Introduction, the starting point is the action functional with target space metric $g$ and 
Wess-Zumino term,
 \bea\label{isomodel2}
 S&=&\int_{\Sigma}\sfrac 12 g_{ij}(X)\dd X^i\wedge \ast \dd X^j
 +\int_{\hat\Sigma} \sfrac 16 H_{ijk}(X)\dd X^i\wedge\dd X^j\wedge\dd X^k~. \eea
The difference to the standard case is that we do not assume \emph{any} global symmetry for $S$. Our guiding principle is 
instead the following: We require the existence of a gauge extension of the action functional $S$ such that it is 
invariant under the \emph{local} symmetry 
\be 
\d_{\epsilon}X^i=\rho^i_a(X)\epsilon^a(\sigma)~,\label{Xgauge}
\ee
without any a priori assumptions on the metric $g$ and the Wess-Zumino term $H$. 

Let us now be more specific. We extend the action $S$ by adding gauge fields taking values in some Lie algebroid $L$ 
$$
L  \overset{\rho}\to  TM~
$$
instead of a Lie algebra. This means that if $e_a$ is a local basis of sections of $L$, then 
$$
A=A^ae_a~.
$$ 
This basis satisfies Eq. \eqref{bracket} and by the map $\rho$ one gets a set of vector fields 
$\rho_a=\rho(e_a)$ which satisfy a similar non-Abelian relation for the Lie bracket with the same structure functions:
\be 
[\rho_a,\rho_b]=C_{ab}^c(X)\rho_c~.
\ee
These gauge fields represent additional would-be dynamical degrees of freedom in the theory. Since we do not wish 
to have extra degrees of freedom in the end, we require that their field strength vanishes. This can be either 
implemented as an additional constraint (see e.g. \cite{Bakas:2016nxt}) or taken care of by adding Lagrange 
multipliers in the action. Here we follow the second approach, which is closer in spirit to the 
traditional duality formulations of 
Buscher \cite{Buscher:1987sk} and Duff \cite{Duff:1989tf}. Moreover, in order to build on the analogy to these 
formulations, we call these additional scalar fields $\widetilde X_a$ (instead of $\eta_a$, which was the notation 
used in Ref. \cite{Chatzistavrakidis:2015lga}). Thus the general form of the candidate 
gauged action we propose is 
 \bea\label{Sgauged}
 S_{\text{gauged}}&=&\int_{\Sigma}\sfrac 12 g_{ij}DX^i\wedge \ast DX^j
 +\int_{\hat\Sigma} \sfrac 16 H_{ijk}\dd X^i\wedge\dd X^j\wedge\dd X^k-\nn\\
 &-& \int_{\Sigma}(\theta_a+\dd\widetilde X_a)\wedge A^a+\int_{\Sigma}\sfrac 12(\iota_{\rho_{[a}}\theta_{b]}+C^c_{ab}(X)\widetilde X_c)A^a\wedge A^b
 -\int_{\Sigma}\omega^a_{bi}\widetilde X_a A^b\wedge DX^i~.
 \eea
 Some comments and clarifications are in order here. 
 First of all, the world sheet covariant derivative $D$ is defined as 
 \be 
 DX^i=\dd X^i-\rho^i_a(X)A^a~,
 \ee 
 as usual. This means that at the level of the kinetic sector we have used minimal coupling of the gauge 
 fields.{\footnote{For a discussion on the generalization to non-minimal kinetic coupling we refer to \cite{strobl}.}}
 However, as already mentioned, minimal coupling does not work for the Wess-Zumino term and this is obvious from the above 
 form of the action, where $A$-dependent terms appear without the involvement of the covariant derivative. 
 Secondly, $\theta_a$ is an 1-form as in the Introduction, namely it can be expanded as $\theta_a=\theta_{ai}(X)\dd X^i$. 
 Finally, the presence of the last term in $S_{\text{gauged}}$ is certainly puzzling at first sight. 
 The involvement of the Lagrange multipliers in this term indicates that it is related to the field strength of the gauge fields. 
 Moreover, it involves some yet undefined parameters $\omega^a_{bi}$. The explanation of these puzzles is one of 
 our main purposes below.
 
 Up to now we have at hand the action $S_{\text{gauged}}$ and the gauge transformation \eqref{Xgauge} of the fields $X^i$, 
 under which we would like it to be invariant. For this to work we have to determine appropriate gauge transformations 
 for the gauge fields $A$ as well as for the scalar fields $\widetilde X_a$. Guided by the corresponding transformation 
 in the standard Lie algebraic and isometric case given by \eqref{AgaugeLie}, we write the gauge 
 transformation for $A$ as 
 \be 
 \d_{\epsilon}A^a=\dd\epsilon^a+C^a_{bc}(X)A^b\epsilon^c+\omega^a_{bi}\epsilon^b DX^i~.
 \ee
 Note that the modification is both in the $X$-dependence of the structure functions $C^a_{bc}$ as well as in the presence of 
 an additional term. This does not yet specify the gauge transformation; it just parametrizes our up to this moment 
 ignorance of the correct transformation such that the action is gauge invariant. Indeed, the parameters 
 $\omega^a_{bi}$ still remain unspecified.{\footnote{In \cite{strobl} an additional 2D-admissible term proportional to 
 the Hodge dual of $DX^i$ was added. Its presence allows for more freedom on the one hand and also yields the 
 relation to generalized geometry more transparent. However here we keep the discussion in its simplest possible form.}} 
 What remains is the gauge transformation for $\widetilde X_a$. This can be determined either the hard way, namely using an 
 arbitrary ansatz, or it can be invoked by extending known results from the standard case (e.g. from \cite{Plauschinn:2014nha}) in a simple way. 
In any case the result is
\be 
\d_{\epsilon}\widetilde X_a=-\iota_{\rho_{(a}}\theta_{b)}\epsilon^b-(C^c_{ab}(X)-\rho^i_{a}\omega^c_{bi})\widetilde X_c\epsilon^b~.
\ee
The cautious reader must have already noticed the repeated appearance of combinations of the structure functions and the parameters 
$\omega^a_{bi}$. In the action \eqref{Sgauged} appears the quantity
\be 
T^c_{ab}=C^c_{ab}(X)-2\rho^i_{[a}\omega^c_{b]i} \label{torsion}
\ee
and thus the transition from the standard case to the non-standard one  goes through a substitution 
$$
C^c_{ab} \to T^c_{ab}(X)~.
$$
This is certainly not an accident. A similar combination was found already in Ref. \cite{Mayer:2009wf}. As explained 
there, although none of $C^a_{bc}$ and $\omega^a_{bi}$ transform as tensors, the combination $T^c_{ab}$ does. 
The geometric meaning of those objects is then the following. The parameters $\omega^a_{bi}$ are coefficients 
of a connection 1-form $\omega^a_b=\omega^a_{bi}\dd X^i$ on the Lie algebroid $L$, namely there is an 
connection $\nabla$ on $L$ such that
\be 
\nabla e_a=\omega^b_a\otimes e_b~.
\ee
This induces also a connection $\nabla_{\rho(\cdot)}$ by means of the map $\rho$ whose torsion $T$ is exactly the one with components 
as in \eqref{torsion}. 
The curvature 2-form of the connection 1-form $\omega^a_b$ may be defined the usual way by the formula
\be 
R^a_b=\dd\omega^a_b+\omega^a_c\w\omega^c_b~.
\ee 
These geometric explanations also shed light to a puzzle encountered previously in 
relation to the last term of the action $S_{\text{gauged}}$. Given the extended gauge transformation of 
the gauge field $A^a$, its field strength should be defined as 
\be 
F^a=\dd A^a+\sfrac 12 C^a_{bc}A^b\w A^c-\omega^a_{bi}A^b\w DX^i~.
\ee 
This is necessary so that $F^a$ has a chance to be covariant. We will not discuss further this field strength 
here, since we do not wish to add dynamics for the gauge fields. However it should now be clear that 
collecting all the terms in $S_{\text{gauged}}$ which contain a Lagrange multiplier, one obtains 
$\widetilde X_a F^a$, as required. In particular, the last term is absolutely necessary for this to 
work.{\footnote{It is mentioned once more that this necessity is only valid in the present approach; one 
could avoid introducing Lagrange multipliers altogether and use constraints instead.}}
 
Let us recapitulate. We have written an action $S_{\text{gauged}}$ which includes the scalar fields 
$X^i$, the gauge fields $A^a$ and the extra scalar fields $\widetilde X_a$, and we know their 
infinitesimal gauge transformations in terms of the structure functions $C^a_{bc}(X)$ and the 
coefficients $\omega^a_{bi}$ of a connection on $L$. Thus it is now a straightforward task to examine 
under which conditions the action is gauge invariant. Before we do so, let us mention that in the 
Lie-algebraic and isometric case there are two invariance conditions that $g$ and $H$ have to satisfy and 
two additional constraints. We will see that in the present case the count of conditions and constraints 
will be the same, with the profit of having milder conditions.

 Indeed, direct variation of the action 
$S_{\text{gauged}}$ reveals that gauge invariance is guaranteed provided that the background fields 
satisfy 
\bea\label{noiso1}
&& {\cal L}_{{\rho}_a}g=\o^b_a\vee\iota_{{\rho}_b}g~,\\
\label{noiso2}
&& \iota_{\rho_a} H=\dd\theta_a+\theta_b\w\omega^b_a -\widetilde X_b R^b_a~,
\eea
where $\vee$ is defined as the symmetric product{\footnote{In completely analogous way to the familiar 
antisymmetric (wedge) product  $\dd X^i\w\dd X^j=\dd X^i\otimes \dd X^j-\dd X^j\otimes \dd X^i$.}} 
 $\dd X^i\vee\dd X^j=\dd X^i\otimes \dd X^j+\dd X^j\otimes \dd X^i$. 
 In component form, Eq. \eqref{noiso1}  is then written as 
 $$({\cal L}_{{\rho}_a}g)_{ij}=\o^b_{ai}\rho_b^kg_{jk}+\o^b_{aj}\rho_b^kg_{ik}~.$$ 
 It is already evident that these conditions allow for non-isometric directions. Indeed 
 for non-vanishing $\omega^a_b$ the Lie derivative of the metric is not any more zero. Thus we have 
 derived the explicit expressions for the right hand sides of Eqs. \eqref{vague1} and 
 \eqref{vague2} advertised in the Introduction. 
  
As anticipated, the above invariance conditions for the background fields are not the end of the story as 
far as gauge invariance is concerned. In direct analogy to the standard case there are two additional constraints, 
\bea\label{constraint1}
&&{\cal L}_{{\rho}_{[a}}\theta_{b]}= C^d_{ab}\theta_d-\iota_{\rho_d}\theta_{[a}\omega^d_{b]}-\iota_{\rho_{[a}}\omega^d_{b]}\theta_d -D^c_{ab} \widetilde X_c~,
\\
\label{constraint2}
&& \sfrac 13 \iota_{\rho_a}\iota_{\rho_b}\iota_{\rho_c}H=\iota_{\rho_{[a}}C^d_{bc]}\theta_d-2\iota_{\rho_{[a}}\omega^d_b \iota_{\rho_{c]}}\theta_d
-2{\tilde D}^d_{abc} \widetilde X_d~,\eea
where we used the following definitions 
\bea
D^e_{ab}&=&\dd C^e_{ab}+C^c_{ab}\omega^e_c+2C^e_{d[a}\omega^d_{b]}
+2\iota_{\rho_d}\omega^e_{[b}\omega^d_{a]}+2{\cal L}_{\rho_{[b}}\omega^e_{a]}
+\iota_{\rho_{[a}}R^e_{b]}~,\\
{\tilde D}^e_{abc}&=&\iota_{\rho_{[a}}\iota_{\rho_b}R^e_{c]}~.
\eea 
In this derivation we used the Jacobi identity \eqref{nc1} for the Lie algebroid $L$. 

It is useful to cross-check that all above expressions are consistent with previously known results 
in the Lie-algebraic and isometric limit. Essentially this is obvious, since in the limits 
$C^a_{bc}(X)\to C^a_{bc}$ and $\omega^a_{bi}\to 0$, the invariance conditions become identical to 
\eqref{isog} and \eqref{isoH}, while the two additional constraints reduce to 
\bea
\label{constraint1red}
&& \eqref{constraint1}\quad  \xrightarrow{\omega^a_b\to 0, ~C^a_{bc}(X)\to C^a_{bc}} \quad {\cal L}_{{\rho}_{[a}}\theta_{b]}= C^d_{ab}\theta_d~,
\\
\label{constraint2red}
&& \eqref{constraint2}\quad   \xrightarrow{\omega^a_b\to 0, ~C^a_{bc}(X)\to C^a_{bc}} \quad\sfrac 13 \iota_{\rho_a}\iota_{\rho_b}\iota_{\rho_c}H=\iota_{\rho_{[a}}C^d_{bc]}\theta_d~,
\eea 
which are identical to the ones found e.g. in \cite{Plauschinn:2014nha}. However, in general the 
conditions we derived are much milder and have the potential to yield gauged actions for vastly more 
initial backgrounds. We discuss whether this potential can be fulfilled later on.

\section{Non-Abelian, non-isometric T-duality}

One of the most direct applications of the type of 2D gauge theories described in the previous section is T-duality 
in string theory. Recall that Buscher's procedure, which leads to the derivation of the widely used T-duality rules for 
background fields, involves a 2D gauge theory with Lagrange multipliers as an intermedium between two string theories 
on different target spaces. The two dual backgrounds are obtained upon integration of different fields in the theory. 
In particular, the integration of the Lagrange multipliers leads back to the original background, while the integration of the gauge fields 
returns a different background whose target space coordinates are essentially the Lagrange multipliers in the action. 
This procedure, be it Abelian \cite{Buscher:1987sk} or non-Abelian \cite{Hull:1989jk,Alvarez:1993qi}, always assumes 
isometric directions from the beginning. 

The essence of our approach here is that isometries are neither an assumption nor a resulting requirement for a 
meaningful, gauge invariant action functional with a local symmetry generated by the vector fields $\rho_a$. 
Thus it is obvious what the next step in our analysis should be. Starting with the action $S_{\text{gauged}}$ 
we should first integrate out the Lagrange multipliers to confirm that the original action is recovered and then we 
should integrate the gauge fields and determine the new action. The calculational details are explained in 
Ref. \cite{Chatzistavrakidis:2015lga}; here we emphasize the final results. 

In order to integrate the Lagrange multipliers $\widetilde X_a$ we vary the action $S_{\text{gauged}}$ with respect to them and 
derive the field equation
\be 
F^a=\dd A^a+\sfrac 12 C^a_{bc}(X) A^b\w A^c-\o^a_{bi}A^b\w DX^i = 0~. 
\ee
This is indeed expected; it means that the non-Abelian gauge fields $A^a$ are pure gauge. We may fix the gauge on-shell, for example 
with the simplest choice being just $A^a=0$, a common choice in the literature (cf. \cite{Rocek:1991ps}). 
Then the action reduces to \eqref{isomodel2}, which is the original action for $g$ and $H$. 

On the other hand, integrating the gauge fields $A^a$ is slightly more involved. First the action is varied with respect to 
them and the resulting field equation is
\be 
\ast \rho^{\ast}_a-\xi_a=G_{ab}\ast A^b+D_{ab}A^b~,\label{Aeom}
\ee
where the following definitions were used:
\bea 
 G_{ab}&=&\rho^i_a g_{ij}\rho^j_b~,\label{Gab}\\
 D_{ab}&=&\iota_{\rho_{[a}}\theta_{b]}+T^c_{ab}\widetilde X_c~,\label{Dab}
 \eea
 and 
 \bea 
 \xi_a&=& \theta_a+\dd\widetilde X_a+\omega^b_a \widetilde X_b~,\\
 \rho^{\ast}_a&=&g_{ij}\rho^i_a\dd X^j~.
 \eea
 Now in order to eliminate the gauge fields from the action, it is required to solve the field equation \eqref{Aeom} 
 for $A^a$. Since this equation involves the differentials $\dd X^i$ and $\dd\widetilde X_a$, as well as their Hodge 
 duals, we anticipate that in general $A^a$ will contain all four corresponding terms. 
 Thus we are naturally led to make the following ansatz:
 \be 
 A^a=M^{ab}\rho^{\ast}_b+N^{ab}\xi_b+P^{ab}\ast\rho^{\ast}_b+Q^{ab}\ast\xi_b~,
 \ee 
 where $M, N, P$ and $Q$ are to be determined. This is essentially the same trick one uses to derive dual models in the 
 standard approaches of Refs. \cite{Buscher:1987sk,Duff:1989tf}. Inserting this ansatz in the relevant field equation, one ends up with a linear system of 
 equations which is solved by 
 \bea 
 Q&=&-(G-DG^{-1}D)^{-1}~,\\
 M&=&-Q~,\nn\\
N&=&-G^{-1}DQ~,\nn\\
P&=&G^{-1}DQ~,\nn
 \eea 
 where $G$ and $D$ are the matrices corresponding to the definitions \eqref{Gab} and \eqref{Dab}. These expressions are 
 not surprising if one compares to similar results in Ref. \cite{Duff:1989tf}.
 
 The last step is to insert the result for $A^a$ in the action $S_{\text{gauged}}$. This then leads to the dual action functional 
 \be 
S_{\text{dual}}=\int_{\S}\left(\sfrac 12 (G-DG^{-1}D)^{ab}e_a\w\ast e_b
-\sfrac 12\big(G^{-1}D(G-DG^{-1}D)^{-1}\big)^{ab}e_a\w e_b \right)+\int_{\hat{\S}} H~,
\ee
where we defined the 1-forms
\be 
e_a=\dd\widetilde X_a+\theta_a-(\omega^b_{ai}\widetilde X_b+(G^{-1}D)^b_a\rho^k_bg_{ki})\dd X^i~.
\ee 
At the \emph{classical} level this is the dual action functional from which a new metric and a new Kalb-Ramond field 
(or 3-form $H$) can be read off. It is observed that the coframe defined by $e_a$ mixes the original scalar fields $X^i$
with the new ones $\widetilde X_a$.{\footnote{This can be related to the embedding of 
the string world sheet in a higher-dimensional geometry. It is useful to note that such embeddings also appear for example in \cite{Mylonas:2012pg,Chatzistavrakidis:2015vka} from a different perspective but still in the context of sigma models.}} Formally this is the case in the standard approach as well, however there it is 
clear that at the end of the day the two sets are disentangled. 
This is not obvious in the present case, however we 
will study some particular examples below to investigate the possibilities and comment accordingly.

\section{Some simple cases}

As mentioned in the Introduction, at this stage one might worry whether \emph{any} non-trivial example 
realizing the above results exists. In other words whether non-zero parameters $\omega^a_{bi}$ can be found 
such that all the invariance conditions and constraints that make $S_{\text{gauged}}$ consistent are 
satisfied. Here we discuss some examples where the procedure indeed works. We already note that they are 
just toy models and they do not correspond to true string backgrounds at the quantum level. Whether 
new dual string backgrounds exist remains an open question that requires a careful analysis which will be 
performed elsewhere. However, certain toy models are useful and often provide valuable hints.    

\paragraph{Abelian$\oplus$Non-isometric.}

First we discuss a simple example where the vector fields $\rho_a$ are Abelian, thus there are no 
structure functions (or, for that matter, even constants) involved; however not all of them generate 
isometries for the metric. 

Let us be more specific. Consider the metric 
\bea\label{mH}
\dd s^2=(\dd x^1)^2+(\dd x^2-x^1\dd x^3)^2 +(\dd x^3)^2.  
\eea
This is the metric of the well-known 3D Heisenberg nilmanifold. This is obtained as the quotient of 
the real 3D Heisenberg group $\text{H}(3;\mathbb{R})$ by its integer counterpart $\text{H}(3;\mathbb{Z})$. 
In more geometric terms it gives rise to a non-trivial 2-torus fibration over a base circle, as it can be 
easily seen by taking the basis of 1-forms 
\be 
e^1=\dd x^1~,\quad e^2=\dd x^2-x^1\dd x^3~,\quad e^3=\dd x^3~,\label{oneforms}
\ee  
and asking it to be globally well-defined. This leads to the identifications 
\be 
(x^1,x^2,x^3) \sim (x^1,x^2+2\pi R,x^3)\sim (x^1,x^2,x^3+2\pi R)\sim (x^1+2\pi R,x^2+2\pi Rx^3,x^3)~,
\ee
where for simplicity we assumed equal radii for the three circles. It is observed that a torus 
$T^2_{(x^2,x^3)}$ is fibered over the circle $S^1_{(x^1)}$. For this reason, this manifold is sometimes called twisted torus in the physics literature. 
The 1-forms \eqref{oneforms} satisfy the Maurer-Cartan equation
\be 
\dd e^2=-C^2_{13}e^1\w e^3~, \quad C^2_{13}=1~.
\ee 
The non-vanishing $C^2_{13}$ is often referred to as geometric flux in the context of string 
compactifications. 

In this example we would like to determine the gauge theory $S_{\text{gauged}}$ and the dual sigma model 
in the case of vanishing Wess-Zumino term, namely $H=0$, and for the choice of vector fields 
$\rho_a=(\partial_1,\partial_2)$, which obviously commute.
 In other words, we would simply like to dualize along the directions 
$x^1$ and $x^2$. Is that possible? First we ask whether the chosen vector fields are Killing. 
The second one, $\rho_2$, indeed is. In fact, choosing to dualize along only this vector field 
produces a dual sigma model with constant Wess-Zumino term $H_{123}$ and target space a 3-torus. 
However the first vector field is not Killing; it satisfies
\bea 
{\cal L}_{\rho_1}g=-\dd x^2\otimes \dd x^3-\dd x^3\otimes \dd x^2+2x^1\dd x^3\otimes \dd x^3~.
\eea
This already indicates what $\omega^a_{bi}$ should be in order to compensate for the non-vanishing 
right hand side of the Lie derivative. The invariance condition \eqref{noiso1} is solved with 
\be 
\omega^2_{13}=-1~.
\ee
Note that all the rest of $\omega^a_{bi}$ are vanishing, and there is no property among indices that 
relates any other to the single non-vanishing component. It is clear that for this solution it holds that 
$R^a_b=0$ and thus the second invariance condition \eqref{noiso2} may be simply solved with the choice 
$\theta_a=0$. Simple inspection of the constraints \eqref{constraint1} and \eqref{constraint2} shows that they 
are satisfied. This means that we indeed have a consistent gauged sigma model with two gauge fields $A^1$ and $A^2$ 
and two Lagrange multipliers $\widetilde X_1$ and $\widetilde X_2$. Integrating out the latter and gauge fixing 
we obtain the sigma model
\be \label{nil}
S=\int_{\S} \sfrac 12 \d_{ab}e^a\w\ast e^b~,
\ee 
which is precisely the purely geometric sigma model with target the Heisenberg nilmanifold. 
However, integrating out the gauge fields through their field equations, 
\bea
A^1&=&\dd X^1+\ast(\dd\widetilde X_1-\widetilde X_2\dd X^3)~,\nn\\
A^2&=&\dd X^2-X^1\dd X^3+\ast\dd\widetilde X_2~,\nn
\eea
we are led to the following 
dual model:
\be 
S_{\text{dual}}=\int_{\S}\left(\sfrac 12 (\dd\widetilde X_1-\widetilde X_2\dd X^3)\w\ast 
(\dd\widetilde X_1-\widetilde X_2\dd X^3)+\sfrac 12 \dd\widetilde X_2\w\ast\dd\widetilde X_2+\sfrac 12 \dd X^3\w\ast\dd X^3 \right)~,
\ee
up to total derivatives.
This comprises a coframe 
\bea 
e_1=\dd\widetilde X_1-\widetilde X_2\dd X^3~,\quad e_2=\dd \widetilde X_2~,\quad e^3=\dd X^3~, 
\eea 
which satisfies 
$$
\dd e_1=-C^2_{13}e_2\w e^3~.
$$
The result has a flavour of self-duality. This is expected though, in view of the fact that 
if one first T-dualizes with respect to $\rho_2$, thus reaching the case of the 3-torus with $H$ flux, 
$\rho_1$ in this intermediate situation is now a Killing vector. Then the T-duality along $\rho_1$ leads 
again to a Heisenberg nilmanifold. At the level of the $B$ field, a gauge transformation is needed in 
this intermediate step. This may be summarized as follows
\bea 
 \xymatrix{ & H_{123}\ar@(ul,ur)^{\d B} \ar[d]^{\text{T}^{\text{iso}}_{\partial_1}}   \\
  \:\, C^2_{13} \ar@/^/[ur]^{\text{T}^{\text{iso}}_{\partial_2}}  \ar[r]^{\text{T}^{\text{non-iso}}_{(\partial_1,\partial_2)}} & C^1_{23}}
\eea
with the diagram being commutative. Thus this example represents a case where the non-isometric approach 
acts as a short-cut in reproducing an otherwise known isometric duality chain. 

\paragraph{Non-Abelian$\oplus$Non-isometric.} 

A more involved example, discussed already in Ref. \cite{Chatzistavrakidis:2015lga}, starts with the same 
manifold $M$ as above, but this time with a choice of non-Abelian vector fields $\rho_a$. The most obvious 
option for such a set are the vector fields dual to the 1-forms $e^a$. 
These are given by 
$$
\rho_a=(\partial_1,\partial_2,\partial_3+x^1\partial_2)~.
$$
Note that the first two are the same as before, but the added one is such that $[\rho_1,\rho_3]=C^2_{13}\rho_2$, 
thus they satisfy the 3D Heisenberg algebra. As before, $\rho_2$ is Killing but $\rho_1$ and $\rho_3$ are not; 
$\rho_3$ satisfies
\be 
{\cal L}_{\rho_3}g=\dd x^1\otimes \dd x^2+\dd x^2\otimes\dd x^1-x^1\dd x^1\otimes \dd x^3-x^1\dd x^3\otimes \dd x^1~.
\ee 
Once more there is no Wess-Zumino term and $\theta_a$ are taken to be zero. Now there are three 
gauge fields $A^1,A^2$ and $A^3$ and three associated Lagrange multipliers. The non-vanishing $\omega^a_{bi}$ 
coefficients that guarantee that all the invariance conditions and constraints are solved now are
{\footnote{A numerical mistake in Ref. \cite{Chatzistavrakidis:2015lga}, which propagated in the ensuing Eqs. \eqref{eomsh}
and \eqref{dualh} is corrected here. This led to a somewhat obscure interpretation of the dual action in \cite{Chatzistavrakidis:2015lga}, which is now 
fully clarified.}}
\be 
 \omega^2_{31}=-\omega^2_{13}=1~.
\ee
Integrating out the Lagrange multipliers one arrives again at the action \eqref{nil}, while the integration of the gauge fields leads first to the field equations 
\bea 
A^1&=&\dd X^1-\sfrac{\widetilde X_2}{1+(\widetilde X_2)^2}\dd\widetilde X_3-\sfrac 1{1+(\widetilde X_2)^2}\ast\dd\widetilde X_1~,\nn\\
A^2&=&\dd X^2-X^1\dd X^3-\ast\dd \widetilde X_2~,\nn\\
A^3&=&\dd X^3+\sfrac{\widetilde X_2}{1+(\widetilde X_2)^2}\dd\widetilde X_1-\sfrac 1{1+(\widetilde X_2)^2}\ast\dd\widetilde X_3~,
\label{eomsh}
\eea 
and upon substitution in $S_{\text{gauged}}$ to the dual action
\bea \label{dualh}
S_{\text{dual}}&=&\frac 12\int_{\S} \left(\dd \widetilde X_2\w\ast \dd \widetilde X_2+
\frac{1}{1+(\widetilde X_2)^2}(\dd\widetilde X_1\w\ast \dd\widetilde X_1+
\dd \widetilde X_3\w\ast \dd\widetilde X_3)+\frac{2\widetilde X_2}{1+(\widetilde X_2)^2}\dd\widetilde X_1\w \dd\widetilde X_3\right)~,\nn\\ 
\eea
up to total derivatives.
This action exhibits a structure identical to that of a T-fold \cite{tfold0,tfold1}, namely a 
non-geometric $Q$ flux background. 
In fact, although in a less expected fashion than in the previous example, we encounter again a 
commutative diagram:
\bea 
 \xymatrix{  H_{123}\ar@(ul,ur)^{\d B} \ar[r]^{\text{T}^{\text{iso}}_{\partial_1}}  &C^1_{23} \ar[d]^{\text{T}^{\text{iso}}_{\partial_3}}   \\
  \:\, C^2_{13} \ar[u]^{\text{T}^{\text{iso}}_{\partial_2}}  \ar[r]^{\text{T}^{\text{non-iso}}_{\rho_a}} & Q^{13}_{2}}
\eea
The commutativity of this diagram is less expected because the Killing vector fields for the isometric 
route are not all the same with the ones in the non-isometric route---in particular the third one is 
$\partial_3$ and $\rho_3=\partial_3+x^1\partial_2$ respectively. Thus in this example we encounter a 
less obvious non-isometric short cut for an isometric duality chain. Of course the real challenge would be 
to perform a T-duality for a case that is completely out of the realm of standard methods. We 
mention such a possibility through the following example.

\paragraph{A note on isometries broken only by the Wess-Zumino term.}

Let us briefly refer to another interesting option arising in the context of non-isometric T-duality. 
Suppose we have a set of Abelian vector fields $\rho_a$, thus $C^a_{bc}=0$, a metric for which these 
vector fields are all Killing, and a non-vanishing Wess-Zumino term given by $H$. 
We encounter now the possibility that although the vector fields generate would-be isometries, these 
isometries are broken by the Wess-Zumino term. Thus, in case $\omega^a_{bi}=0$ we face a serious 
obstacle: although the invariance conditions \eqref{noiso1} and \eqref{noiso2} are satisfied (at 
least for some choice of $\theta_a$), the constraint \eqref{constraint2}, reduced now to \eqref{constraint2red}, 
cannot be satisfied. This was also noticed in Ref. \cite{Plauschinn:2014nha} and it is related to the 
problem of finding a triple T-dual of the torus with $H$ flux. Although we are not going to solve this problem 
here, we now indicate a possible direction for its potential solution. 
Although the Lie derivative of the metric is zero, \eqref{noiso1} does not mean that $\omega^a_{bi}$ 
has to vanish. Instead it just means that 
$$ 
\rho_b^k\omega^b_{a(i}g_{j)k}=0~,
$$
which is much milder. For example, for the simple case of $\rho_a=\d_a^i\partial_i$ and $g_{ij}=\d_{ij}$, it 
reduces to 
$$
\omega_{a(ij)}=0~,
$$
which is solved by any set of coefficients antisymmetric in the indices involved in the above 
symmetrization. The point is that the previously lethal constraint now reads as in Eq. \eqref{constraint2} 
and its right hand side is not any more necessarily zero. This allows for the possibility of solving 
the constraints for such cases too, previously impossible. We plan to report on this issue in a future 
publication.

\section{Take-home messages}

The main messages of this work may be summarized as follows
\begin{itemize}
 \item Given background fields $g$ and $B$, there exist classically consistent gauged 2D sigma models of maps $X=(X^i):\S\to M$
  whose gauge symmetry is generated by vector fields that do not necessarily generate isometries.
 \item These gauged sigma models can act as intermediate gauge theories to study candidate T-dual string backgrounds beyond the 
 realm of isometry. 
 \item Non-trivial toy models that realize such non-Abelian and non-isometric T-duality do exist.
\end{itemize}

Although these results are certainly encouraging it is equally useful to keep in mind the limitations of the approach
presented here. Some of these are the following: ($\a$) The analysis is limited to the classical 
level. It is not yet clear whether our results survive quantization, ($\beta$) in relation to the above, we were only 
able at this stage to verify that the approach works non-trivially in toy models and not in true, conformal string 
backgrounds, ($\g$) the dilaton was simply ignored, ($\d$) the fully worked-out examples where our approach 
indeed works are so far just short cuts for results that can be obtained by the standard method. A merit test for our 
approach would be, for instance, the proper derivation of the triple T-dual of a torus threaded by $H$ flux.
Further work is required in order to answer questions posed by these remarks. 

\paragraph{Acknowledgements.} Collaboration with A. Deser, L. Jonke and T. Strobl is gratefully acknowledged. 
Helpful discussions with V. Penas and L. Romano are kindly appreciated too. I am grateful to M. Bugden for pointing out two typos 
in the first version. Finally, I would like to express heartfelt thanks to 
George Zoupanos and Ifigenia Moraiti, the pillars of the activities of the European Institute for Sciences and their 
Applications (EISA) in Corfu.

\end{document}